\documentclass[conference]{IEEEtran}

\usepackage{times,amsmath,amssymb,amsfonts,amsthm,graphicx}
\usepackage{cite,color}
\usepackage{bm}
\usepackage{balance}
\usepackage{url}
\usepackage{subfigure}
\usepackage{booktabs}

\makeatletter
\newcommand*{\rom}[1]{\expandafter\@slowromancap\romannumeral #1@}
\makeatother

\def\thetabf{\boldsymbol \theta}

\def\ebf{{\bf e}}
\def\fbf{{\bf f}}
\def\gbf{{\bf g}}
\def\hbf{{\bf h}}

\def\nbf{{\bf n}}

\def\pbf{{\bf p}}

\def\sbf{{\bf s}}

\def\ubf{{\bf u}}
\def\vbf{{\bf v}}
\def\wbf{{\bf w}}

\def\zbf{{\bf z}}

\def\Ibf{{\bf I}}

\def\Bc{{\cal B}}

\def\Kc{{\cal K}}

\def\Mc{{\cal M}}

\def\Pc{{\cal P}}

\def\Sc{{\cal S}}

\def\ie{{\it i.e.,\ \/}}
\def\nn{\nonumber}

\def\Re{\mathfrak{R}\mathfrak{e}}
\def\Im{\mathfrak{I}\mathfrak{m}}

\def\mae{{\mathbb{E}}}

\theoremstyle{definition}

\newtheorem{assumption}{Assumption}
\newtheorem{proposition}{Proposition}

\newenvironment{mylist}%
{\begin{list}{}%
    {%
      \setlength{\itemindent}{0pt}%
      \setlength{\leftmargin}{10pt}%
      \setlength{\parsep}{\parskip}
      \setlength{\labelsep}{10pt}
      \setlength{\itemsep}{2pt}}}%
  {\end{list}}

\def\BibTeX{{\rm B\kern-.05em{\sc i\kern-.025em b}\kern-.08em
    T\kern-.1667em\lower.7ex\hbox{E}\kern-.125emX}}

\begin{document}

\title{\huge Uplink Over-the-Air Aggregation for Multi-Model\\ Wireless Federated Learning}

\author{ Chong Zhang$^{\star}$, Min Dong$^{\dagger}$, Ben Liang$^{\star}$, Ali Afana$^{\ddagger}$, Yahia Ahmed$^{\ddagger}$\\
\normalsize $^{\star}$Dept. of Electrical and Computer Engineering, University of Toronto, Canada, $^{\ddagger}$Ericsson Canada, Canada   \\
$^{\dagger}$Dept. of Electrical, Computer and Software Engineering, Ontario Tech University, Canada\thanks{This work was funded in part by Ericsson Canada and by the Natural Sciences and Engineering Research Council (NSERC) of Canada.}
}%

\maketitle

\begin{abstract}
We propose an uplink over-the-air aggregation (OAA) method for wireless federated learning (FL) that simultaneously trains multiple models. To maximize the multi-model training convergence rate, we derive an upper bound on the optimality gap of the global model update, and then, formulate an uplink joint transmit-receive beamforming optimization problem to minimize this upper bound. We solve this problem using the block coordinate descent approach, which admits low-complexity closed-form updates. Simulation results show that our proposed multi-model FL with fast OAA substantially outperforms sequentially training multiple models under the conventional single-model approach.
\end{abstract}

\section{Introduction}
\label{sec:intro}

Federated learning (FL) \cite{Mcmahan&etal:2017} is a widely recognized method for multiple devices to collaboratively train machine learning models. However, FL in the wireless environment, usually with a base station (BS) taking the role of a parameter server, suffers from degraded performance due to limited wireless resources and signal distortion. This necessitates efficient communication design to effectively support wireless FL.

Most existing works on wireless FL have focused on training only a single model  \cite{Amiri&etal:TWC2022,Yang&etal:TWC2020,Zhang&Tao:TWC2021,Sun&etal:JSAC2022,Guo&etal:JSAC2022,Wang&etal:JSAC2022b,Zhang&etal:arxiv2023}.
Various design schemes have been proposed to improve the communication efficiency of wireless FL, including transmission design of the downlink \cite{Amiri&etal:TWC2022},
uplink \cite{Yang&etal:TWC2020,Zhang&Tao:TWC2021,Sun&etal:JSAC2022},
and combined downlink-uplink \cite{Guo&etal:JSAC2022,Wang&etal:JSAC2022b,Zhang&etal:arxiv2023}.
However, in practice a system often needs to train multiple models.
Directly using the conventional single-model FL schemes, to train the models sequentially one at a time,
can cause substantial latency.

Simultaneously training multiple models in FL has recently been considered in \cite{Bhuyan&etal:2023,Zhang&etal:ICASSP2024}.
Under error-free communication, it was shown in \cite{Bhuyan&etal:2023} that multi-model FL can substantially improve the training convergence rate. Later, considering noisy downlink and uplink wireless channels, \cite{Zhang&etal:ICASSP2024} proposed a multi-group multicast beamforming method to facilitate the downlink transmission of global models from the BS to the devices. However, \cite{Zhang&etal:ICASSP2024} used the conventional orthogonal multiple access design for uplink model aggregation, which can consume large bandwidth and incur high latency as the number of devices becomes large.
While over-the-air aggregation (OAA) has recently become popular for uplink design in single-model FL due to its bandwidth efficiency over orthogonal schemes \cite{Yang&etal:TWC2020,Zhang&Tao:TWC2021,Sun&etal:JSAC2022},
it has not been considered in multi-model FL, due to the substantial design challenges from additional inter-model interference and high computational complexity.

In this paper, we propose a computationally efficient uplink OAA method for multi-model wireless FL.
Aiming to maximize the FL convergence rate, we derive an upper bound on the optimality gap of the FL global model
update, capturing the impact of noisy transmission and inter-model interference.
We then show that the minimization of this upper bound leads to a joint transmit-receive beamforming design to minimize the sum of inverse received SINRs subject to some power budget at the BS and devices.
We solve this problem using block coordinate descent (BCD) and derive closed-form solutions to each subproblem, leading to a low-complexity  design. Simulation under typical wireless network settings shows that the proposed multi-model FL design with fast OAA substantially outperforms the conventional single-model approach that sequentially trains one model at a time.

\allowdisplaybreaks
\section{System Model and Problem Statement} \label{sec:system_prob}
\subsection{Multi-Model FL System}\label{sec:fl_model}
\vspace*{-.25em}
We consider an FL system consisting of a server and $K$
worker devices that collaboratively train $M$ ML models.
Let  $\Kc_{\text{tot}} \triangleq \{1, \ldots, K\}$ denote the total set of devices and  $\Mc \triangleq \{1, \ldots, M\}$  the set of models.
Let $\thetabf_m\in\mathbb{R}^{D_m}$ be the parameter vector of model $m$, which has $D_m$ parameters.

Each device $k$ holds local training datasets for all $M$ models,
with the dataset for model $m$ being $\Sc_m^{k} \triangleq \{(\sbf_{m,i}^k,v_{m,i}^k): 1 \le i \le S_m^k\}$, where $\sbf_{m,i}^k\in\mathbb{R}^{b} $ is the $i$-th data feature vector and $v_{m,i}^k$ is the corresponding label.
The local training loss function representing the training error
at device $k$ for model $m$ is defined as
$F_m^{k}(\thetabf_m)=\frac{1}{S_m^k}\sum_{i=1}^{S_m^k} L_m(\thetabf_m;\sbf_{m,i}^k,v_{m,i}^k)$,
where $L_m(\cdot)$ is the sample-wise training loss for model $m$.
 The global training loss function for model $m$ is a weighted average of
$F_m^{k}(\thetabf_m)$'s, given by
$F_m(\thetabf_m) = \frac{1}{\sum_{k=1}^{K}S_m^{k}}\sum_{k=1}^{K}S_m^{k}F_m^{k}(\thetabf_m)$.
The learning objective is to find optimal $\thetabf_m^\star$ that minimizes $F_m(\thetabf_m)$ for
each model $m\in\Mc$.

\begin{figure}[t]
\centering
\hspace{-1em}\includegraphics[scale=0.37]{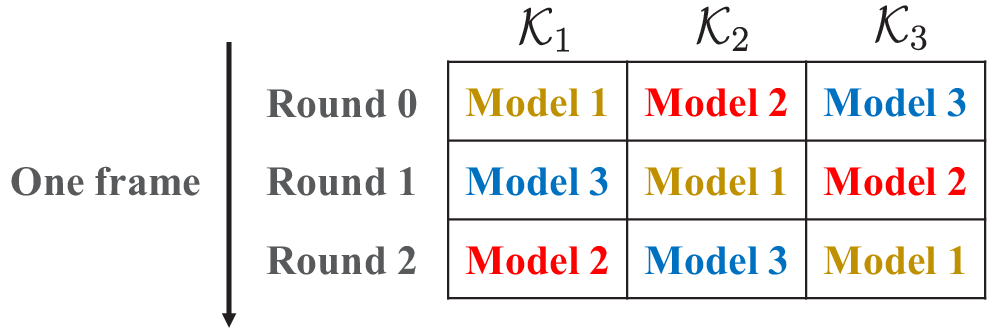}
\vspace*{-.7em}\caption{Device-model  round robin scheduling for $M=3$ models.}\vspace*{-1.5em}
\label{fig1:rr}
\end{figure}

For multi-model FL, we consider the $K$ devices train the $M$ models simultaneously, and the model updates are exchanged with the server via multiple rounds of downlink-uplink wireless communication. In each communication round, each model is trained by a subset of devices.
For simplicity, we assume $K/M \in \mathbb{N}$. We consider the round robin scheduling approach for device-model assignment \cite{Bhuyan&etal:2023,Zhang&etal:ICASSP2024}.   Specifically, we define a frame consisting of $M$ communication rounds.
At the beginning of each frame, the $K$ devices are randomly partitioned into $M$ equal-sized groups, denoted by $\Kc_1,\ldots,\Kc_M$. These device groups remain unchanged within a frame.
For each communication round $t$ within the frame, each  device group $i$ is assigned to train  model $\hat{m}(i, t)$, given by
\vspace*{-.3em}
\begin{align}
\hat{m}(i, t) = [(M + i - [t \,\,\text{mod}\,\, M] - 1) \,\,\text{mod}\,\, M] + 1.\label{eq_rr_schedule}
\end{align}
\\[-1.5em]
Fig.~\ref{fig1:rr} shows an example of the round-robin device-model assignment within a frame of three communication rounds for $M=3$ models.

The iterative multi-model FL training procedure in round $t$,
which is in frame $n= \lfloor t/M\rfloor$,
 is as follows:\vspace*{-.2em}
\begin{mylist}
\item\emph{Downlink broadcast}: The server  broadcasts the current $M$ global model parameter vectors $\thetabf_{m,t}$'s to their respective assigned device group.

\item \emph{Local model update}: Device $k\in\Kc_i$ performs local training of its assigned model $\hat{m}(i, t)$  using the corresponding local dataset $\Sc_{\hat{m}(i, t)}^{k}$.
Suppose $\hat{m}(i, t)=\mu$. Device $k$ divides $\Sc^k_{\mu}$
into mini-batches, and applies the standard mini-batch stochastic
gradient descent (SGD) algorithm with
$J$ iterations to generate the updated local model based on the
received version of the global model $\hat{\thetabf}^k_{\mu,t}$.
In particular, let $\thetabf^{k,\tau}_{\mu,t} $ denote the local model update by device $k\in \Kc_i$ at iteration $\tau \in \{0,\ldots,J-1\}$, with
$\thetabf^{k,0}_{\mu,t} = \hat{\thetabf}^k_{\mu,t}$, and let $\Bc^{k,\tau}_{\mu,t}
\subseteq \Sc^k_{\mu}$  denote the mini-batch used  at iteration  $\tau$.
The local model update is given by
\begin{align}
        \thetabf^{k,\tau+1}_{\mu,t}
        = \thetabf^{k,\tau}_{\mu,t} - \frac{\eta_n}{|\Bc^{k,\tau}_{\mu,t}|}\sum_{(\sbf,v)
        \in\Bc^{k,\tau}_{\mu,t}}\nabla L_{\mu}(\thetabf^{k,\tau}_{\mu,t}; \sbf,v) \label{SGD}
\end{align}
where $\eta_n$ is the learning rate in frame $n$,
and $\nabla L_{\mu}$ is the gradient of the sample-wise training loss function  for model $\mu$ w.r.t. $\thetabf^{k,\tau}_{\mu,t}$.

\item\emph{Uplink aggregation}: The $K$ devices send their updated local models $\thetabf^{k,J}_{m,t}$'s to the server using the uplink transmission.
The server aggregates $\thetabf^{k,J}_{m,t}$, $k\in\Kc_i$, received from device group $i$ to generate the global model $\thetabf_{m,t+1}$
for each $m\in\Mc$ for the next round $t+1$.
\end{mylist}

\subsection{Wireless Communication Model}
We consider a practical wireless communication system where
the server is hosted by a BS. Assume the BS is equipped with $N$ antennas, and each device has a single antenna.

We assume downlink broadcast of $M$ models to their respective device groups uses orthogonal channels among groups. The BS uses the multicast beamforming
technique \cite{Sidiropoulos&etal:TSP2006,Dong&Wang:TSP2020} to
send the model update  $\thetabf_{\hat{m}(i, t),t}$
to its assigned device group $i$. Device $k$ in group $i$ then obtains an estimate of $\thetabf_{\hat{m}(i, t),t}$ \cite{Zhang&etal:ICASSP2024}:
\begin{align}
        \hat{\thetabf}^{k}_{\hat{m}(i, t),t}= \thetabf_{\hat{m}(i, t),t}  + \nbf^{\text{dl}}_{k,t}
        \label{dl_device_signal}
\end{align}
where $\nbf^{\text{dl}}_{k,t}\sim \mathcal{N}({\bf 0}, \sigma^2_{\text{d}}\Ibf)$ is the post-processed receiver noise
due to the noisy downlink channel.

For uplink transmission and local model aggregation, we consider OAA to conserve system bandwidth.
In particular, the $K$ devices send their local model updates $\thetabf^{k,J}_{m,t}$'s
to the BS simultaneously over a common uplink channel.
The BS uses receive beamforming to aggregate
the local models $\thetabf^{k,J}_{\hat{m}(i, t),t}$, $k\in\Kc_i$, received from device group $i$, for $i=1,\ldots,M$.
Due to the analog nature of OAA, the devices must send the values of $\thetabf^{k,J}_{m,t}$'s directly under their transmit power budget.

In this paper, we focus on the uplink OAA design via joint transmit-receiver beamforming, aiming to maximize the learning performance of multi-model wireless FL in terms of the training convergence rate.
Note that the downlink received models are noisy versions of $\thetabf_{m,t}$'s due to the noisy wireless channel, while the uplink received models are also distorted versions of $\thetabf^{k,J}_{m,t}$'s due to the noisy wireless channels and the inter-group interference. These errors  further propagate in the
model updates over the subsequent communication rounds, degrading the learning performance. Thus, an effective uplink OAA design must capture such errors generated in the complex interaction between learning and communication.

\section{Uplink OAA for Multi-Model FL}\label{sec:FL_alg}

\subsection{Uplink Aggregation Framework}\label{subsec:ul_aggre}
We propose an uplink aggregation framework where the devices simultaneously send the multiple local model updates $\thetabf^{k,J}_{m,t}$'s to
the BS via the common uplink channel.
Recall that $\thetabf^{k,J}_{m,t} \in \mathbb{R}^{D_m}$.
For efficient transmission,  we convert $\thetabf^{k,J}_{m,t}$ into an equivalent complex vector representation  $\tilde{\thetabf}^{k,J}_{m,t}$,
whose real and imaginary parts respectively contain the first and second halves of the elements in $\thetabf^{k,J}_{m,t}$.
That is,
$\tilde{\thetabf}^{k,J}_{m,t} = \tilde{\thetabf}^{k,J\text{re}}_{m,t} + j\tilde{\thetabf}^{k,J\text{im}}_{m,t}\in\mathbb{C}^{\frac{D_{m}}{2}}$, where
$\tilde{\thetabf}^{k,J\text{re}}_{m,t}$ contains the first $D_m\over 2$ elements in  $\thetabf^{k,J}_{m,t}$ and   $\tilde{\thetabf}^{k,J\text{im}}_{m,t}$  contains the rest $D_m\over 2$ elements.

We assume the uplink channels remain unchanged within one frame.
Let $\hbf_{k,n}\in\mathbb{C}^{N}$ denote the channel from device $k$ to the BS in frame $n$,
which is assumed known perfectly at the BS.
Let $a_{k,n}\in\mathbb{C}$ be the transmit beamforming weight at device $k$ in frame $n$ for sending its local model update.
Let $D_{\text{max}}\triangleq\max_{m\in\Mc} D_m$.
Under perfect synchronization, all $K$ devices  simultaneously send their respective normalized complex model updates, $\frac{\tilde{\thetabf}^{k,J}_{m,t}}{\|\tilde{\thetabf}^{k,J}_{m,t}\|}$'s, to the BS using $\frac{D_{\text{max}}}{2}$ channel uses in round $t$.
For model $\hat{m}(i,t)=m$ with  $D_{m} < D_{\text{max}}$,  a random   position is set for all $k\in \Kc_i$. Each device $k\in \Kc_i$ uses this position for $\tilde{\thetabf}^{k,J}_{m,t}$ within $\frac{D_{\text{max}}}{2}$ channel uses and applies zero padding to the rest of positions. Thus, the equivalent signal vector
at this device $k$ is $\bar{\thetabf}^{k,J}_{m,t} = [{\bf{0}}^H, (\tilde{\thetabf}^{k,J}_{m,t})^H, {\bf{0}}^H]^H$ with length $D_{\text{max}} \over 2$.
The received signal vector $\vbf_{l,t}\in \mathbb{C}^{N}$ at the BS in channel use $l$ is given by\vspace*{-.5em}
\begin{align}
\vbf_{l,t} = \sum_{i=1}^{M}\sum_{k\in\Kc_i}\hbf_{k,n}a_{k,n}\frac{\bar{\theta}^{k,J}_{ml,t}}{\|\tilde{\thetabf}^{k,J}_{m,t}\|} + \ubf^{\text{ul}}_{l,t}, \quad l=1,\ldots,\frac{D_\text{max}}{2} \nn
\end{align}
where $\ubf^{\text{ul}}_{l,t}\sim \mathcal{CN}({\bf 0}, \sigma^2_{\text{u}}\Ibf)$  is the receiver noise vector with i.i.d. zero mean and variance $\sigma^2_{\text{u}}$.

The BS applies  receive beamforming to $\vbf_{l,t}$'s for over-the-air aggregation of $\tilde{\thetabf}^{k,J}_{m,t}$, $k\in\Kc_i$, from each group $i$.
Let $\wbf^{\text{ul}}_{i,n}\in\mathbb{C}^{N}$ be the unit-norm receive beamforming vector at the BS for group $i$ in frame $n$, with $\|\wbf^{\text{ul}}_{i,n}\|^2 = 1$.
For device $k\in\Kc_i$, and assume $\hat{m}(i,t)=m$, its effective  channel   after  the BS receive beamforming is given by $\alpha^{\text{ul}}_{k,t} \triangleq \frac{(\wbf^{\text{ul}}_{i,n})^H\hbf_{k,n}a_{k,n}}{\|\tilde{\thetabf}^{k,J}_{m,t}\|}$. Thus, the corresponding post-processed received signal vector for  $\tilde{\thetabf}^{k,J}_{m,t}$ over the $\frac{D_m}{2}$ channel uses  is given by
\begin{align}
\zbf_{m,t}\!=\!\!\! \sum_{k\in\Kc_i}\!\!\!\alpha^{\text{ul}}_{k,t}\tilde{\thetabf}^{k,J}_{m,t}
\!\!+\!\! \sum_{j\neq i}\sum_{q\in\Kc_j}\!\!(\!\wbf^{\text{ul}}_{i,n}\!)^H\!\hbf_{q,n}a_{q,n}\!\frac{\bar{\thetabf}^{'q,J}_{\hat{m}(j, t),t}}{\|\tilde{\thetabf}^{q,J}_{\hat{m}(j, t),t}\|}\!\! +\! \nbf^{\text{ul}}_{m,t}. \nn
\end{align}
where $\bar{\thetabf}^{'q,J}_{\hat{m}(j, t),t}\in\mathbb{C}^{\frac{D_m}{2}} $ is the portion of other (zero-padded) model  $\bar{\thetabf}^{q,J}_{\hat{m}(j, t),t}$ sent by device $q\in \Kc_j$ that aligns with the location of $\tilde{\thetabf}^{k,J}_{m,t}$ in $\bar{\thetabf}^{k,J}_{m,t}$, and $\nbf^{\text{ul}}_{m,t}$ is the post-processed receiver noise with  the $l$-th element being $(\wbf^{\text{ul}}_{i,n})^H\ubf^{\text{ul}}_{l,t}$, for  $l=1,\ldots,\frac{D_m}{2}$.

We consider  uplink joint transmit-receive beamforming, where $\{a_{k,n}\}_{k\in\Kc_i}$ and  $\wbf^{\text{ul}}_{i,n}$ are designed jointly for each device group $i$ in frame $n$. For  OAA to be effective, the local models $\tilde{\thetabf}^{k,J}_{m,t}$'s need to be added coherently. Thus, the transmit and receive beamforming design should ensure that the resulting effective channels  $\alpha^{\text{ul}}_{k,t}$'s, for $k\in\Kc_i$ in group $i$, are phase aligned.
Thus, we set the transmit beamforming weight  $a_{k,n} =\sqrt{p_{k,n}}  \frac{\hbf_{k,n}^{H}\wbf^{\text{ul}}_{i,n}}{|\hbf^{H}_{k,n}\wbf^{\text{ul}}_{i,n}|}$, for $k\in \Kc_i$,
where $p_{k,n}$ is the transmit power of this device. The effective channels of all devices in group $i$ are then phase aligned to $0$ after receive beamforming as
\begin{align*}
\alpha^{\text{ul}}_{k,t} = \frac{(\wbf^{\text{ul}}_{i,n})^H\hbf_{k,n}a_{k,n}}{\|\tilde{\thetabf}^{k,J}_{m,t}\|} = \frac{\sqrt{p_{k,n}}|\hbf^{H}_{k,n}\wbf^{\text{ul}}_{i,n}|}{\|\tilde{\thetabf}^{k,J}_{m,t}\|}, \ k \in \Kc_i.
\end{align*}
Each device is subject to the transmit power budget.
Let  $D_{\text{max}}P^{\text{ul}}_k$ be the total transmit power budget for sending the entire normalized local model in $\frac{D_{\text{max}}}{2}$ channel uses at device $k$, where $2P^{\text{ul}}_k$ denotes the average transmit power budget per channel use
for sending two elements. Then, for transmitting $\frac{\tilde{\thetabf}^{k,J}_{m,t}}{\|\tilde{\thetabf}^{k,J}_{m,t}\|}$, we have the transmit power constraint $p_{k,n} \le D_{\text{max}}P^{\text{ul}}_k$.

After receive beamforming, the BS further scales $\zbf_{m,t}$ to obtain the complex equivalent global model update
$\tilde{\thetabf}_{m,t+1}$ for the next round $t+1$, where $m=\hat{m}(i,t)$:
\begin{align}
& \tilde{\thetabf}_{m,t+1} = \frac{\zbf_{m,t}}{\sum_{k\in\Kc_i}\alpha^{\text{ul}}_{k,t}} = \sum_{k\in\Kc_i}\rho_{k,t}\tilde{\thetabf}^{k,J}_{m,t}
+ \tilde{\nbf}^{\text{ul}}_{m,t} \nn\\
& + \!\frac{1}{\sum_{k\in\Kc_i}\!\alpha^{\text{ul}}_{k,t}}\!\sum_{j\neq i}\!\sum_{q\in\Kc_j}  \!\!\!\frac{\hbf^H_{q,n}\!\wbf^{\text{ul}}_{j,n}(\wbf^{\text{ul}}_{i,n})^H\hbf_{q,n}}{|\hbf^H_{q,n}\wbf^{\text{ul}}_{j,n}|}\! \cdot\! \frac{\sqrt{p_{q,n}}\bar{\thetabf}^{'q,J}_{\hat{m}(j, t),t}}{\|\tilde{\thetabf}^{q,J}_{\hat{m}(j, t),t}\|}   \label{eq_global_update_0}
\end{align}
where $\rho_{k,t}  \triangleq \frac{\alpha^{\text{ul}}_{k,t}}{\sum_{q\in\Kc_i}\alpha^{\text{ul}}_{q,t}}$ is the weight with  $\sum_{k\in\Kc_i}\rho_{k,t} = 1$,
 and $\tilde{\nbf}^{\text{ul}}_{m,t}  \triangleq  \frac{\nbf^{\text{ul}}_{m,t}}{\sum_{k\in\Kc_i}\alpha^{\text{ul}}_{k,t}}$ is
 the post-processed receiver noise at the BS. The weight $\rho_{k,t}$ represents the uplink processing effect including the device transmission and BS receiver processing.

Let $\tilde{\thetabf}_{m,t}$ and $\tilde{\nbf}^{\text{dl}}_{k,t}$ denote the equivalent complex representations of
$\thetabf_{m,t}$ and $\nbf^{\text{dl}}_{k,t}$ in \eqref{dl_device_signal}, respectively, for $m=\hat{m}(i,t)$.
For local model update  in \eqref{SGD},
$\Delta\tilde{\thetabf}^{k}_{m,t} \triangleq \tilde{\thetabf}^{k,J}_{m,t} - \tilde{\thetabf}^{k,0}_{m,t}$ is the equivalent complex representation of the local model difference after the local training at device $k\in\Kc_i$ in round $t$.
Using \eqref{dl_device_signal} and \eqref{eq_global_update_0}, we obtain the global model update $\tilde{\thetabf}_{m,t+1}$ from    $\tilde{\thetabf}_{m,t}$ as
\begin{align}
& \tilde{\thetabf}_{m,t+1} = \tilde{\thetabf}_{m,t} + \sum_{k\in\Kc_i}\rho_{k,t}\Delta\tilde{\thetabf}^{k}_{m,t}
+ \sum_{k\in\Kc_i}\rho_{k,t}\tilde{\nbf}^{\text{dl}}_{k,t} +  \tilde{\nbf}^{\text{ul}}_{m,t} \nn\\
& + \!\frac{1}{\sum_{k\in\Kc_i}\!\alpha^{\text{ul}}_{k,t}}\!\sum_{j\neq i}\!\sum_{q\in\Kc_j}  \!\!\!\frac{\hbf^H_{q,n}\wbf^{\text{ul}}_{j,n}(\wbf^{\text{ul}}_{i,n})^H\hbf_{q,n}}{|\hbf^H_{q,n}\wbf^{\text{ul}}_{j,n}|}\! \cdot\! \frac{\sqrt{p_{q,n}}\bar{\thetabf}^{'q,J}_{\hat{m}(j, t),t}}{\|\tilde{\thetabf}^{q,J}_{\hat{m}(j, t),t}\|}
\label{eq_global_update}
\end{align}
Finally, the real-valued global model update $\thetabf_{m,t+1}$ can be recovered from its complex version as
$\thetabf_{m,t+1}\!=[\Re{\{\tilde{\thetabf}_{m,t+1}\}^T}\!, \Im{\{\tilde{\thetabf}_{m,t+1}\}^T}]^T$.

\subsection{Multi-Model FL Convergence Analysis under Uplink OAA}
\label{sec:upper_bound}

Our objective is to design uplink joint transmit-receive beamforming to minimize
the maximum expected optimality gap to $\thetabf_m^\star$ among all $M$ models after $S$ frames, subject to the transmitter power budget.
In particular,  let $\Sc\triangleq \{0,\ldots,S-1\}$. Let $\pbf_n\triangleq[\pbf^T_{1,n}, \ldots, \pbf^T_{M,n}]^T$, with $\pbf_{i,n}\in\mathbb{R}^{\frac{K}{M}}$ being the power vector containing  $p_{k,n}$, $k\in\Kc_i$ of group $i$ in frame $n$. Also, let
$\wbf^{\text{ul}}_n\triangleq[(\wbf^{\text{ul}}_{1,n})^H, \ldots, (\wbf^{\text{ul}}_{M,n})^H]^H\in\mathbb{C}^{MN}$
 denote the BS receive beamforming vector for all $M$ groups in frame $n$.
Our optimization problem can be formulated as
\begin{align}
\Pc_{o}: \,\,\min_{\{\wbf^{\text{ul}}_n,\pbf_n\}^{S-1}_{n=0}} \,\, &\max_{m\in\Mc} \, \mae[\|\thetabf_{m,SM}- \thetabf_m^\star\|^2] \nn\\
\text{s.t.} \ \ &  p_{k,n} \le D_{\text{max}}P^{\text{ul}}_k, \ k\in \Kc_{\text{tot}},\,\, n\in\Sc,\nn\\
& \|\wbf^{\text{ul}}_{i,n}\|^2 = 1, \quad i\in \Mc,\,\, n\in\Sc \nn
\end{align}
where $\mae[\cdot]$ is  taken w.r.t.\ receiver noise and
mini-batch local data samples  at each device.
Problem $\Pc_{o}$ is a  stochastic optimization problem with a min-max objective. To tackle this challenging problem, we develop a more tractable upper bound on $\mae[\|\thetabf_{m,SM}- \thetabf_m^\star\|^2]$ by analyzing the convergence rate of the global model update.

We make the following  assumptions on
the local loss functions, the local model updates, and the divergence of
the global and local loss gradients.
They are commonly used in the convergence
analysis of FL training \cite{Amiri&etal:TWC2022,Guo&etal:JSAC2022,Bhuyan&etal:2023}.
\vspace*{-.3em}
\begin{assumption}\label{assump_smooth}
The local loss function $F_m^{k}(\cdot)$ is $L$-smooth and $\lambda$-strongly convex,
$\forall m\in\Mc$, $\forall k\in\Kc_{\text{tot}}$.
\end{assumption}
\begin{assumption}\label{assump_bound_model}
Bounded local model parameters:
$\|\tilde{\thetabf}^{k,J}_{m,t}\|^2 \leq r$,
for some $r>0$, $\forall m\in\Mc$, $\forall k\in\Kc_{\text{tot}}$, $\forall t$.

\end{assumption}
\begin{assumption}\label{assump_bound_diff}
Bounded  gradient divergence of loss functions: $\mae[\| \nabla F_m(\thetabf_{m,t})-\sum_{k=1}^{K}c_{k}\nabla F^k_{m}(\thetabf^{k,\tau}_{m,t})  \|^2] \leq \phi$
and
 $\mae[\|\nabla F^k_{m}(\thetabf^{k,\tau}_{m,t}) -\nabla F^k_{m}(\thetabf^{k,\tau}_{m,t}, \Bc^{k,\tau}_{m,t})
 \|^2] \leq \delta$
 for some $\phi \ge 0$, $\delta\geq 0$, $0\leq c_k \leq 1$,
$\forall m\in\Mc$, $\forall k\in\Kc_{\text{tot}}$, $\forall \tau$, $\forall t$.
\end{assumption}
\vspace*{-.3em}
Based on  \eqref{eq_global_update}, we first obtain the per-model global update  equation over frames.
Let  device group   $\hat{i}$ be the group that trains model $m$ in communication round $t$ in frame $n. $ The device-model assignment  between $\hat{i}$ and $m$ is given in   \eqref{eq_rr_schedule}. Summing both sides of \eqref{eq_global_update} over $M$ rounds in  frame $n$, and subtracting the optimal $\tilde{\thetabf}_m^\star$ (complex version of  ${\thetabf}_m^\star$) from both sides,
we obtain
\vspace*{-.5em}
\begin{align}
\tilde{\thetabf}_{m,(n+1)M} \!-\! \tilde{\thetabf}_m^\star \!\! = \!\tilde{\thetabf}_{m,nM}\! -\! \tilde{\thetabf}_m^\star
+\!\!\!\!\!\!\! \sum_{t=nM}^{(n+1)M-1}\!\!\!\!\!\!\sum_{k\in\Kc_{\hat{i}}}\!\rho_{k,t}\Delta\tilde{\thetabf}^{k}_{m,t}
\!+ \!\tilde{\ebf}_{m,n} \nn
\end{align}
where $\tilde{\ebf}_{m,n}$ is the accumulated error term  in \eqref{eq_global_update}  over $M$ rounds in frame $n$, given by
\begin{align}
& \tilde{\ebf}_{m,n} \triangleq \sum_{t=nM}^{(n+1)M-1}\sum_{k\in\Kc_{\hat{i}}}\rho_{k,t}\tilde{\nbf}^{\text{dl}}_{k,t} +  \sum_{t=nM}^{(n+1)M-1}\tilde{\nbf}^{\text{ul}}_{m,t}  \nn\\
&+\!\!\! \sum_{t=nM}^{(n+1)M-1}\!\!\!\!\!\sum_{j\neq \hat{i}}\sum_{q\in\Kc_j}  \!\!\frac{\hbf^H_{q,n}\wbf^{\text{ul}}_{j,n}(\wbf^{\text{ul}}_{\hat{i},n})^H\hbf_{q,n}}{|\hbf^H_{q,n}\wbf^{\text{ul}}_{j,n}|\sum_{k\in\Kc_{\hat{i}}}\!\alpha^{\text{ul}}_{k,t}} \cdot \frac{\sqrt{p_{q,n}}\bar{\thetabf}^{'q,J}_{\hat{m}(j, t),t}}{\|\tilde{\thetabf}^{q,J}_{\hat{m}(j, t),t}\|}.\nn
\end{align}
By Assumption~\ref{assump_bound_model},
we can further bound $\mae[\|\tilde{\ebf}_{m,n}\|^2]$ as
\begin{align}
\mae\big[\|\tilde{\ebf}_{m,n}\|^2\big]\! & \!\leq rMK\!\sum_{i=1}^{M}\!\frac{\sum_{j\neq i}\sum_{q\in\Kc_j}p_{q,n} |\hbf^{H}_{q,n}\wbf^{\text{ul}}_{i,n}|^2 +\tilde{\sigma}^2_{\text{u}}}{(\sum_{k\in\Kc_i}\sqrt{p_{k,n}}|\hbf^{H}_{k,n}\wbf^{\text{ul}}_{i,n}|)^2}  \nn\\
&\qquad + 2K\tilde{\sigma}^2_{\text{d}} \label{eq_bound_error}
\end{align}
where $\tilde{\sigma}^2_{\text{d}} \triangleq \sigma^2_{\text{d}}D_{\text{max}}/2$ and $\tilde{\sigma}^2_{\text{u}} \triangleq \sigma^2_{\text{u}}D_{\text{max}}/2$.

Using the above, we obtain an upper bound on $\mae[\|\thetabf_{m,SM}- \thetabf_m^\star\|^2]$ below. The proof is omitted due to space limitation.
\begin{proposition}\label{thm:convergence}
Under  Assumptions~\ref{assump_smooth}--\ref{assump_bound_diff} and for  $\eta_n<\frac{1}{\lambda}$, $\forall n$, the expected optimality gap after $S$ frames is bounded by
\vspace*{-.5em}
\begin{align}
\!\mae[\|\thetabf_{m,SM}- \thetabf_m^\star\|^2]\! \leq & \Gamma_m\!\prod_{n=0}^{S-1}\!G_n  \!+\Lambda\!+\! \sum_{n=0}^{S-2}\!H(\wbf^{\text{ul}}_n,\pbf_n)\!\!\!\prod_{s=n+1}^{S-1}\!\!\!G_s\nn\\
&   + H(\wbf^{\text{ul}}_{S-1},\pbf_{S-1}), \quad m\in\Mc
\label{eq_thm1}
\end{align}
where $\Gamma_m \triangleq \mae[\| \thetabf_{m,0} - \thetabf_m^\star\|^2]$,  $G_n \triangleq 4(1-\eta_n\lambda)^{2JM}$, $\Lambda \triangleq \sum_{n=0}^{S-2}C_{n} \big(\!\prod_{s=n+1}^{S-1}\!G_s\big)+ C_{S-1}$ with $C_{n} \!\triangleq 4\eta^2_{n}J^2
(M^2\phi + K^2\delta)+8K\tilde{\sigma}^2_{\text{d}}$, and
\vspace*{-.7em}
\begin{align}
H(\wbf^{\text{ul}}_n,\pbf_n)\! & \triangleq\! 4rMK\!\sum_{i=1}^{M}\!\frac{\sum_{j\neq i}\sum_{q\in\Kc_j}p_{q,n} |\hbf^{H}_{q,n}\wbf^{\text{ul}}_{i,n}|^2\! +\!\tilde{\sigma}^2_{\text{u}}}{(\sum_{k\in\Kc_i}\sqrt{p_{k,n}}|\hbf^{H}_{k,n}\wbf^{\text{ul}}_{i,n}|)^2}. \nn
\end{align}
\end{proposition}

\subsection{Uplink Joint Transmit-Receive Beamforming Design}
\label{sec:ul_joint_bf}

We now replace the objective function in $\Pc_{o}$
with  the more tractable upper bound in \eqref{eq_thm1}.
Omitting the first two constant terms  in \eqref{eq_thm1} that do not depend on the beamforming design,  we arrive at the following equivalent optimization problem:
\vspace*{-.5em}
\begin{align}
\Pc_{1}: \,\,  \min_{\{\wbf^{\text{ul}}_n,\pbf_n\}^{S-1}_{n=0}} \,\, &  \sum_{n=0}^{S-2}\!H(\wbf^{\text{ul}}_n,\pbf_n)\!\!\!\prod_{s=n+1}^{S-1}\!\!G_s  + H(\wbf^{\text{ul}}_{S-1},\pbf_{S-1}) \nn\\
\text{s.t.} \ \ & p_{k,n} \le D_{\text{max}}P^{\text{ul}}_k, \quad k\in \Kc_{\text{tot}},\,\, n\in\Sc,\nn\\
& \|\wbf^{\text{ul}}_{i,n}\|^2 = 1, \quad i\in \Mc,\,\, n\in\Sc. \nn
\end{align}
By Proposition~\ref{thm:convergence}, for $\eta_n<\frac{1}{\lambda}$, we have $G_n > 0$, $\forall n$. Thus,  $\Pc_{1}$ can be decomposed into $S$ subproblems,
one for each frame $n$, given by
\vspace*{-.5em}
\begin{align}
\Pc_{2,n}: \,\,\min_{\wbf^{\text{ul}}_n,\pbf_n} \,\, & \sum_{i=1}^{M}\frac{\sum_{j\neq i}\sum_{q\in\Kc_j}p_{q,n} |\hbf^{H}_{q,n}\wbf^{\text{ul}}_{i,n}|^2 +\tilde{\sigma}^2_{\text{u}}}{(\sum_{k\in\Kc_i}\sqrt{p_{k,n}}|\hbf^{H}_{k,n}\wbf^{\text{ul}}_{i,n}|)^2} \nn\\
\text{s.t.} \ \ & p_{k,n} \le D_{\text{max}}P^{\text{ul}}_k, \quad k\in \Kc_{\text{tot}},\nn\\
& \|\wbf^{\text{ul}}_{i,n}\|^2 = 1, \quad i\in \Mc. \nn
\end{align}

Problem $\Pc_{2,n}$ is a multi-user joint uplink  transmit power allocation and receive beamforming problem with a complicated objective function of
$\{\wbf^{\text{ul}}_n,\pbf_n\}$. To make the problem amenable for a solution, we consider an upper bound of the objective function.
Let $\fbf_{k,n} \triangleq \hbf_{k,n}/\tilde{\sigma}_{\text{u}}$.
Since  $(\sum_{k\in\Kc_i}\sqrt{p_{k,n}}|\fbf^{H}_{k,n}\wbf^{\text{ul}}_{i,n}|)^2 \ge \sum_{k\in\Kc_i}p_{k,n}|\fbf^{H}_{k,n}\wbf^{\text{ul}}_{i,n}|^2$, we replace the objective function in $\Pc_{2,n}$ by an upper bound and arrive at the following problem:
\begin{align}
\Pc_{3,n}: \,\,\min_{\wbf^{\text{ul}}_n,\pbf_n} \,\, &
\sum_{i=1}^{M}\frac{\sum_{j\neq i}\sum_{q\in\Kc_j}p_{q,n} |\fbf^{H}_{q,n}\wbf^{\text{ul}}_{i,n}|^2 +1}{\sum_{k\in\Kc_i}p_{k,n}|\fbf^{H}_{k,n}\wbf^{\text{ul}}_{i,n}|^2}\nn\\
\text{s.t.} \ \ & p_{k,n} \le D_{\text{max}}P^{\text{ul}}_k, \quad k\in \Kc_{\text{tot}},\nn\\
& \|\wbf^{\text{ul}}_{i,n}\|^2 = 1, \quad i\in \Mc. \nn
\end{align}

We note that in the objective function, the $i$th term in the summation is the inverse of SINR for the aggregated local models received from group $i$, and the objective function is the sum of inverse SINRs of all $M$ groups. For this jointly non-convex problem  $\Pc_{3,n}$, we apply BCD  to solve it, \ie alternatingly updates the BS receive beamforming $\wbf^{\text{ul}}_n$
and the  device powers in $\pbf_n$.
The two subproblems are given below:

\emph{1) Updating} $\wbf^{\text{ul}}_n$: Given $\pbf_n$,  $\Pc_{3,n}$ can be further decomposed into $M$ subproblems, one for each beamformer  $\wbf^{\text{ul}}_{i,n}$ as
\begin{align}
\Pc^{\text{wsub}}_{3,n,i}: \,\,\min_{\wbf^{\text{ul}}_{i,n}} \,\, &
\frac{(\wbf^{\text{ul}}_{i,n})^H\left(\sum_{j\neq i}\sum_{q\in\Kc_j}
p_{q,n}\fbf_{q,n}\fbf^H_{q,n} + \Ibf\right)\wbf^{\text{ul}}_{i,n}}
{(\wbf^{\text{ul}}_{i,n})^H\left(\sum_{k\in\Kc_i}p_{k,n}\fbf_{k,n}\fbf^H_{k,n}\right)\wbf^{\text{ul}}_{i,n}}\nn\\
\text{s.t.} \ \ & \|\wbf^{\text{ul}}_{i,n}\|^2 = 1, \nn
\end{align}
which is a generalized eigenvalue problem.
The optimal solution $\wbf^{\text{ul}}_{i,n}$ can be obtained in closed-form, which is the  generalized eigenvector corresponding to
the smallest generalized eigenvalue.
 We omit the detail due to space limitation.

\emph{2) Updating $\pbf_{n}$}:
Let $\gbf_{ij,n}$ be a $\frac{K}{M} \times 1$ vector containing $\{g_{iq,n}\triangleq |\fbf^{H}_{q,n}\wbf^{\text{ul}}_{i,n}|^2$,  $q\in\Kc_{j}\}$ of group $j\in\Mc$. Given $\{\wbf^{\text{ul}}_{i,n}\}$, we can rewrite $\Pc_{3,n}$ as
\begin{align}
\Pc^{\text{psub}}_{3,n}: \,\,\min_{\{\pbf_{i,n}\}^{M}_{i=1}} \,\, &
\sum_{i=1}^{M}\frac{\sum_{j\neq i}\gbf^T_{ij,n}\pbf_{j,n} +1}{\gbf^T_{ii,n}\pbf_{i,n}}\nn\\
\text{s.t.} \ \ & p_{k,n} \le D_{\text{max}}P^{\text{ul}}_k,
\quad k\in \Kc_{\text{tot}}. \nn
\end{align}
We propose to update $\pbf_{1,n},\ldots,\pbf_{M,n}$ sequentially
using BCD.
Given  $\pbf_{j,n}$, $\forall j\in\Mc, j\neq i$,  $\Pc^{\text{psub}}_{3,n}$
is convex w.r.t. $\pbf_{i,n}$ for group $i$,
for which the optimal $\pbf_{i,n}$ can be obtained in closed-form via the KKT conditions.
Specifically, let $\bar{\pbf}^{\text{ul}}_{i}$ be the vector containing
$\{D_{\text{max}}P^{\text{ul}}_k$, $k\in\Kc_{i}\}$ of group $i$.
Let $a^{\text{min}}_{i,n} \triangleq \min_{k\in\Kc_i}\Big(\frac{\sum_{j\neq i} \gbf^T_{ij,n}\pbf_{j,n} +1}{ \sum_{j\neq i}\frac{g_{jk,n}} {\gbf^T_{jj,n}\pbf_{j,n}}}g_{ik,n}\Big)^{1/2}$ and let $k'\in \Kc_i$ be the corresponding index that achieves $a^{\text{min}}_{i,n}$.
Thus, the optimal $\pbf_{i,n}$ is given by
\begin{align}
p_{k,n} =
\begin{cases}
D_{\text{max}}P^{\text{ul}}_{k}     & \text{for~} k\in\Kc_i, k\neq k' \\
D_{\text{max}}P^{\text{ul}}_{k'} - \frac{\displaystyle \left[\gbf^T_{ii,n}\bar{\pbf}^{\text{ul}}_{i}-a^{\text{min}}_{i,n}\right]^+}{\displaystyle g_{ik',n}}     & \text{for~} k=k' \nn
\end{cases}
\end{align}
where $[a]^+ =\max\{a,0\}$.
Thus,  $\pbf_{i,n}$ is updated sequentially using the above solution.
\section{Simulation Results}
\label{sec:simulations}
We consider image classification under the current cellular  system setting with  $10$ MHz bandwidth and  $2$~GHz carrier frequency.
The maximum transmit  powers at the BS and devices are $47~\text{dBm}$ and $23~\text{dBm}$, respectively. We assume the devices use $1$ MHz bandwidth for uplink transmission.
Each channel is generated as $\hbf_{k,t} = \sqrt{G_{k}}\bar{\hbf}_{k,t}$, where
$\bar{\hbf}_{k,t}\sim\mathcal{CN}({\bf{0}},{\bf{I}})$, and the path gain
$G_{k} [\text{dB}] = -136.3-35\log_{10}d_k - \psi_k$,
with the BS-device distance  $d_k\in(0.02, 0.5)$   in kilometers and the shadowing random variable  $\psi_k$ having the standard deviation  $8~\text{dB}$.
 Noise power spectral density is $ -174~\text{dBm/Hz}$,
and we assume noise figure $N_F=8$~dB and $2$ dB at the device and  BS receivers, respectively. We set $N = 64$ and $K=12$.

We use the MNIST dataset for the multi-model training and testing. It consists of $60,000$ training samples and $10,000$ test samples.
We train three types of convolutional neural networks:
i) \textbf{Model A}:  an $8\times3\times3$ ReLU convolutional layer,
a $2\times2$ max pooling layer, and a softmax output layer
with $13,610$ parameters.
ii) \textbf{Model B}:  a $6\times4\times4$ ReLU convolutional layer,
a $2\times2$ max pooling layer, a ReLU fully-connected layer with $22$ units, and a softmax output layer
with $19,362$ parameters.
ii) \textbf{Model C}:  an $8\times3\times3$ ReLU convolutional layer,
a $2\times2$ max pooling layer, a ReLU fully-connected layer with $20$ units, and a softmax output layer
with $27,350$ parameters.
We use the $10,000$ test samples to measure the test accuracy of each global model update $\thetabf_{m,t}$ at round $t$.
The training samples are randomly and evenly distributed  across devices, with the local dataset size $S_{k} = {60,000}/{K}$ samples at device $k$.
For the local training using SGD, we set $J=100$, the mini-batch size $|\Bc^{k\tau}_{m,t}|={600}/{K}, \forall k,\tau,m,t$, and the learning rate $\eta_n=0.1$, $\forall n$.
All results are obtained by taking the current best test accuracy and averaging over $20$ channel realizations.

\begin{figure}[t]
\centering
\includegraphics[width=.52\columnwidth]{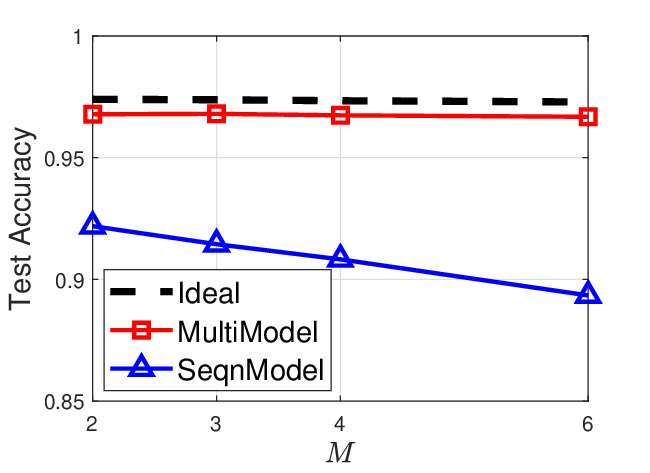}\hspace*{-.3em}\includegraphics[width=.52\columnwidth]{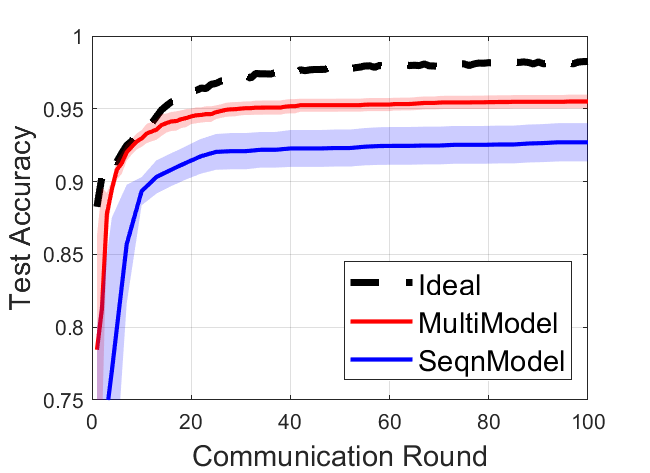}
\includegraphics[width=.52\columnwidth]{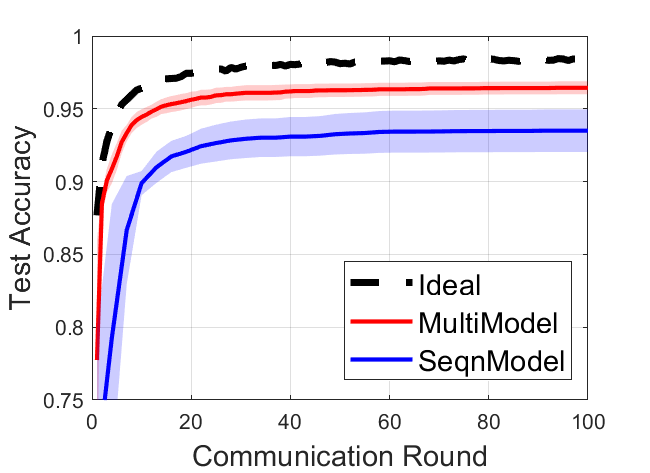}\hspace*{-.3em}\includegraphics[width=.52\columnwidth]{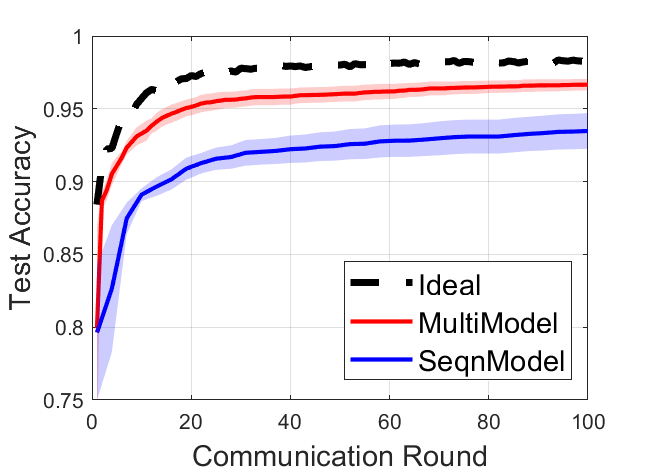}
\vspace*{-2em}\caption{\footnotesize Top Left: Test accuracy vs. $M$ (Model A).
Rest of figures: Test accuracy vs. communication round $t$: Top Right -- Model A; Bottom Left -- Model B; Bottom Right -- Model C ($90\%$ confidence intervals are shown).}
\label{Fig:Accuracy_N64_K12} \vspace*{-1.5em}
\end{figure}

We denote our proposed method as MultiModel. We also consider  two schemes for comparison:
i) \textbf{Ideal}: Multi-model FL\ via  \eqref{eq_global_update} with error-free downlink and uplink. It serves as a performance upper bound for all schemes.
ii) \textbf{SeqnModel}: Sequentially train each model using the single-model FL  with all $K$ devices by the uplink beamforming scheme that maximizes the aggregated SNR  provided in [8].
Fig.~\ref{Fig:Accuracy_N64_K12}-Top Left shows the test accuracy vs.\   $M$ models, after $T=30$   rounds, where all models are from Model A.  We see that MultiModel substantially outperforms the sequential model training  for all $M$ values.
We also consider  mixed model types. We set $M=3$, one from each of Models A, B, and C. Fig.~\ref{Fig:Accuracy_N64_K12}-Top Right, Bottom Left, and Bottom Right show the test accuracy
over round $t$ for Models A, B, and C, respectively.
We see that  MultiModel outperforms the sequential training using the single-model-based scheme for all models.

\section{Conclusion}
This paper considers uplink transmission design for multi-model wireless FL. We design uplink beamforming for sending multiple models simultaneously to the BS via OAA  to maximize the FL training performance. We utilize an upper bound on the optimality gap of the global multi-model update to formulate the joint uplink transmit-receive beamforming problem and apply BCD to solve it with closed-form iteration updates.
Simulation results demonstrate substantial  performance advantage of the proposed multi-model scheme over the conventional single-model sequential training.

\balance
\bibliographystyle{IEEEbib}
\bibliography{Refs}

\end{document}